\documentclass{naturefig}
\usepackage{amsmath}
\usepackage{amssymb}
\usepackage{graphicx}
\usepackage{color}
\usepackage{siunitx}
\usepackage{pdfpages}

\title{Quantum distillation and confinement of vacancies in a doublon sea}

\author{Lin Xia$^{1}$, Laura A. Zundel$^{1}$, Juan Carrasquilla$^{1,2}$, Aaron Reinhard$^{1}$,
Joshua M. Wilson$^{1}$, Marcos Rigol$^{1}$, and David S. Weiss$^{1}$}

\begin{document}

\date{\today}
\maketitle
\begin{affiliations}
\item Department of Physics, The Pennsylvania State University,
  University Park, PA 16802, USA
\item Perimeter Institute for Theoretical Physics, Waterloo, Ontario, Canada N2L 2Y5
\end{affiliations}

\begin{abstract}
Ultracold atomic gases have revolutionized the study of non-equilibrium dynamics in quantum
many-body systems. Many counterintuitive non-equilibrium effects have been observed, such as
suppressed thermalization in a one-dimensional (1D)
gas,\cite{kinoshita_wenger_06} the formation of repulsive self-bound dimers,\cite{winkler_thalhammer_06} 
and identical behaviors for attractive and repulsive
interactions.\cite{ronzheimer_schreiber_13} Here, we observe the expansion of a bundle of ultracold
1D Bose gases in a flat-bottomed optical lattice potential. By combining in situ measurements with
photoassociation,\cite{theis_thalhammer_04,kinoshita_wenger_04} we follow the spatial dynamics
of singly, doubly, and triply occupied lattice sites. The
system sheds interaction energy by dissolving some doublons and triplons. Some singlons quantum
distill out of the doublon center,\cite{fabian_manmana_09,muth_petrosyan_12} while others remain
confined.\cite{muth_petrosyan_12} Our Gutzwiller mean-field model captures these experimental
features in a physically clear way. These experiments
might be used to study thermalization in systems with particle
losses\cite{makotyn_klauss_14} or the evolution of quantum
entanglement,\cite{cheneau2012,Daley_Zoller_12} or if applied to fermions, to prepare very
low entropy states.\cite{fabian_manmana_09}
\end{abstract}

%%%%%%%%%%%%%%%%%%%%%%%%%%%%%%%%%%%%%%%%%%%%%%%%%%%%%%%%%%%%%%%%%%%%%%%%%
%%%%%%%%%%%%%%%%%%%%%%%%%%%%%%%%%%%%%%%%%%%%%%%%%%%%%%%%%%%%%%%%%%%%%%%%%

\baselineskip24pt

\newpage

%%%%%%%%%%%%%%%%%%%%%%%%%%%%%%%%%%%%%%%%%%%%%%%%%%%%%%%%%%%%%%%%%%%%%%%%%

% Introduction

\begin{figure}
  \begin{center}
    \includegraphics[width=0.9\columnwidth]{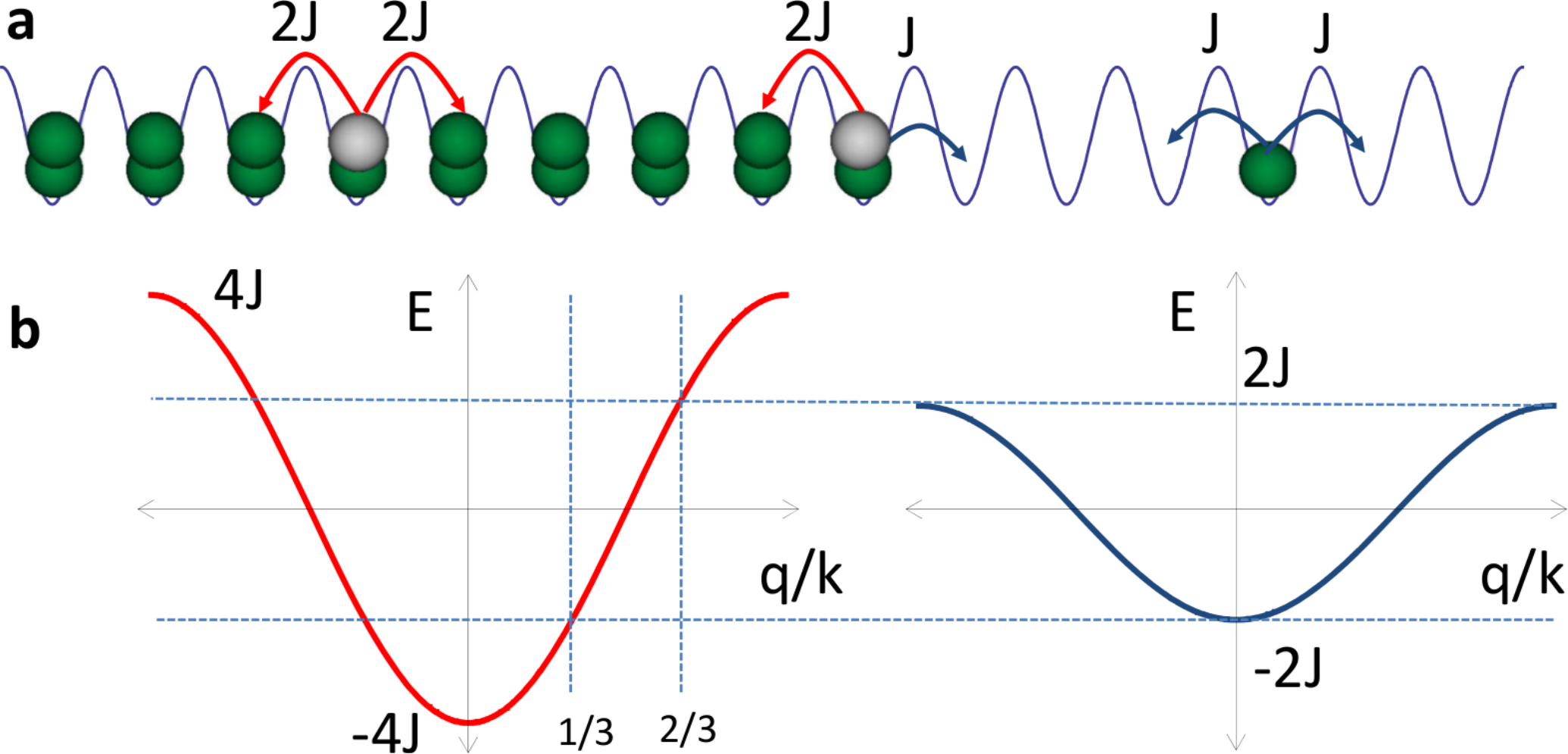}
  \end{center}
  \vspace{-0.4cm}
  \caption{\textbf{a}. Boson quantum distillation cartoon. A singlon among
  doublons acts as a vacancy (shown in grey), which tunnels through the doublon sea at $2J$.
  A singlon in the empty lattice tunnels at $J$. A vacancy at the edge of the doublon sea is
  the same as a singlon at the edge of the empty lattice.  It is important to note that,
  at our low temperature and our lattice depths, atoms are delocalized. Each 1D tube is occupied
  by a superposition of many distributions like the one pictured, generally including sites
  with higher occupancies and empty sites, as well as different doublon edge positions.
  \textbf{b}. Ground energy bands within the doublon sea (on the right) and in the empty
  lattice (on the left). Only atoms with quasimomentum between 1/3 and 2/3 of the band edge
  can tunnel from the doublon sea into the empty lattice. Outside that range transmission
  into the empty lattice does not conserve energy. This limitation on quantum distillation
  is absent for fermions.\label{fig:1}}
\end{figure}

%%%%%%%%%%%%%%%%%%%%%%%%%%%%%%%%%%%%%%%%%%%%%%%%%%%%%%%%%%%%%%%%%%%%%%%%%
Quantum distillation is a previously unobserved phenomenon in which atoms at singly occupied lattice sites (singlons) escape the central region of an untrapped lattice gas, leaving doubly occupied lattice sites (doublons) behind  (see Figure~\ref{fig:1}\textbf{a}).\cite{fabian_manmana_09} It depends on one readily achievable condition, that there be an energy mismatch that prevents isolated doublons from disintegrating into two singlons.
Singlons in a sea of doublons can be understood as vacancies . The tunneling rate of a singlon in an empty lattice is $J$, while the bosonic vacancy tunneling rate is 2$J$, since the vacancy moves when either of the adjacent doublon's atoms tunnels. When bosonic vacancies reach a doublon sea edge only those with intermediate energies can transmit into the empty lattice while conserving energy,\cite{muth_petrosyan_12} as illustrated in Figure~\ref{fig:1}\textbf{b}. Vacancies with quasimomenta outside that limited range reflect from the edge, confining them in the doublon sea. When a singlon does exit, it purifies and shrinks the doublon sea by one lattice site. It has been hypothesized that collisions of vacancies with triplons can thermalize the vacancies, possibly transferring them into transmissible quasimomentum states.\cite{muth_petrosyan_12} In contrast, fermionic vacancies have the same tunneling energy as singlons in the empty lattice, so they always pass through the edge of the doublon sea.

A previous non-equilibrium experiment with bosons in flat 1D lattices mostly focused on initial
single-atom number states in a deep lattice, and studied the expansion dynamics after a quench of the
on-site interaction energy.\cite{ronzheimer_schreiber_13} Such a quench leaves the many-body
wavefunction out of local equilibrium. In contrast, we start our experiments with
trapped, superfluid, 1D Bose gases in lattices with an average of between one and two atoms per site,
and we quench by suddenly removing the trap. This is a fundamentally different quench, a
geometric quench, after which the many-body wavefunction is still locally in equilibrium. Geometric
quenches have been theoretically shown to lead to remarkable universal phenomena, such as quasi-condensation
at finite momenta,\cite{rigol_muramatsu_04,rodriguez_manmana_06} dynamical fermionization of the
Tonks-Girardeau gas,\cite{rigol_muramatsu_05,minguzzi_gangardt_05} and identical expansions of
bosons and fermions.\cite{vidmar_langer_13} Our experimentally observed expansion dynamics are qualitatively
reproduced by a Gutzwiller mean-field calculation (see Methods).
By theoretically and experimentally studying the spatial evolution
of site occupancy, we obtain a straightforward physical interpretation of the dynamics. It exhibits
clear signatures of quantum distillation\cite{fabian_manmana_09} and
confinement\cite{muth_petrosyan_12,jreissaty_carrasquilla_13} of vacancies in the doublon sea.

%%%%%%%%%%%%%%%%%%%%%%%%%%%%%%%%%%%%%%%%%%%%%%%%%%%%%%%%%%%%%%%%%%%%%%%%%%

% Experimental setup

%%%%%%%%%%%%%%%%%%%%%%%%%%%%%%%%%%%%%%%%%%%%%%%%%%%%%%%%%%%%%%%%%%%%%%%%%%

In our experiment, Bose condensed $^{87}$Rb atoms in a crossed dipole trap are slowly loaded
into an array of 1D tubes formed by a blue-detuned 2D optical lattice (wavevector $k$=2$\pi$/773 nm),
with a superposed axial optical
lattice of variable depth $V_0$ and a red-detuned crossed dipole trap for overall confinement (see Methods).
For the density and $V_0$ used in this work, the initial ground states are predominantly
superfluid.\cite{rigol_batrouni_09} Since each lattice site starts with a superposition of number states,
pictures like those in Figure~\ref{fig:1}\textbf{a} represent one of many distributions whose coherent
sum is the state of the system. As we will see, most of the qualitative behavior of what are thus
delocalized singlons and doublons can be understood using localized pictures, leavened by the understanding
that each picture represents only a small piece of the overall wavefunction.

At $t_\text{ev}=0$, we suddenly lower the depth of the crossed-dipole
trap, leaving enough power to cancel the residual anti-trap due to the 2D lattice beams over a range
of $\sim$\SI{160}{\micro\metre} (see Methods). We observe the subsequent spatial evolution in three
ways (referred to as M1, M2, and M3), which together allow us to separately determine the spatial
evolution of the probability distributions of singlons, doublons, and atoms at more highly occupied
sites. In M1, we measure all the atoms after a given $t_\text{ev}$ by switching to a 27$E_\text{rec}$ deep
1D axial lattice (where the recoil energy $E_\text{rec}=\hbar^2k^2/2M$, and $M$ is the Rb mass) and
allowing the atoms to expand radially so that the density is low enough for absorption imaging.
Information about the transverse distribution among tubes is lost, but the axial distribution is
preserved, with a resolution of $\sim$\SI{3}{\micro\metre}. In M2, at $t_\text{ev}$ we suddenly switch
to a 27$E_\text{rec}$ lattice in each of three directions and turn on a photoassociation pulse for
1.5\,ms,\cite{theis_thalhammer_04,kinoshita_wenger_04} which is long enough to eliminate all doublons,
2/3 of the triplon atoms, and most of the atoms from sites with higher occupancies. The axial
distribution measurement is then made as in M1. In M3, at $t_\text{ev}$ we suddenly switch to the
27$E_\text{rec}$ 3D lattice and wait for 50\,ms, which is long enough for three-body inelastic
collisions\cite{burt_ghrist_97,tolra_ohara_04} to empty the triplon sites and most
atoms from more highly occupied sites. The axial distribution is then measured as in M1.

%%%%%%%%%%%%%%%%%%%%%%%%%%%%%%%%%%%%%%%%%%%%%%%%%%%%%%%%%%%%%%%%%%%%%%%%%
% FIGURE 1
%%%%%%%%%%%%%%%%%%%%%%%%%%%%%%%%%%%%%%%%%%%%%%%%%%%%%%%%%%%%%%%%%%%%%%%%%

%%%%%%%%%%%%%%%%%%%%%%%%%%%%%%%%%%%%%%%%%%%%%%%%%%%%%%%%%%%%%%%%%%%%%%%%%
% END FIGURE 1
%%%%%%%%%%%%%%%%%%%%%%%%%%%%%%%%%%%%%%%%%%%%%%%%%%%%%%%%%%%%%%%%%%%%%%%%%

%%%%%%%%%%%%%%%%%%%%%%%%%%%%%%%%%%%%%%%%%%%%%%%%%%%%%%%%%%%%%%%%%%%%%%%%%%

% Results

%%%%%%%%%%%%%%%%%%%%%%%%%%%%%%%%%%%%%%%%%%%%%%%%%%%%%%%%%%%%%%%%%%%%%%%%%%

\begin{figure}
  \begin{center}
    \includegraphics[width=1\columnwidth]{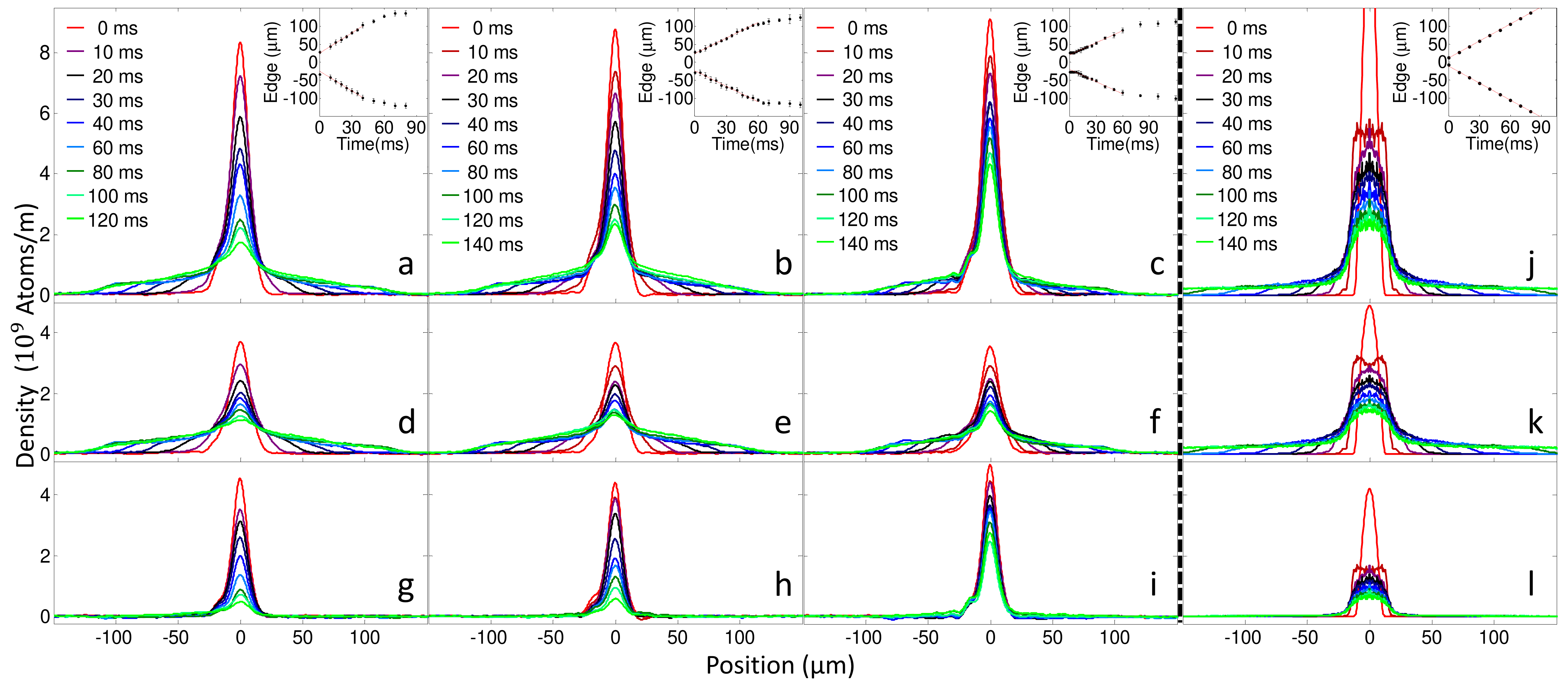}
  \end{center}
  \vspace{-0.8cm}
  \caption{Spatial dynamics in the flat lattice. \textbf{a-c}. M1 measurements of the total atom
  distributions at successive $t_\text{ev}$ at $V_0$=3$E_\text{rec}$, 4$E_\text{rec}$, and 5$E_\text{rec}$,
  respectively. Insets: The location of the distribution edges vs.~time. The measured velocity values for
  $V_0=3E_\text{rec}$, $4E_\text{rec}$, and $5E_\text{rec}$ are $1.9\pm0.18$, $1.44\pm0.02$, and
  $1.12\pm0.03$ mm/s, respectively, slightly slower than $v_\text{max}$ of 2.1, 1.6, and 1.2 mm/s.
  \textbf{d-f}. The distributions of singlons at successive $t_\text{ev}$ at $V_0$=3$E_\text{rec}$,
  4$E_\text{rec}$, and 5$E_\text{rec}$, respectively. These are derived from combining the raw
  measurements according to the formula M2-(M1-M3)/3. \textbf{g-i}. The distributions of doublons
  at successive $t_\text{ev}$ at $V_0$=3$E_\text{rec}$, 4$E_\text{rec}$, and 5$E_\text{rec}$,
  respectively. These are derived from combining the raw measurements according to the formula
  M1-M2-2(M1-M3)/3. Note that all the experimental plots (\textbf{a-i}) have the same vertical scale.
  For a discussion of the small asymmetry in these figures, see Methods. \textbf{j-l.} Gutzwiller
  mean-field theory results for $V_0=4E_\text{rec}$ for the distribution of all atoms, singlons, and
  doublons, respectively. We choose the theory initial conditions so that the fraction of singlons
  matches the experiment (see Methods). The initial fraction of triplons and higher exceed what is
  seen in the experiment, and the initial cloud length is smaller. The qualitative behavior seen in
  the theory  (and in the experiment) is robust to the initial conditions. All but the least occupied
  $\sim20\%$ of tubes behave qualitatively like the average over all tubes, but with less doublon
  dissolution and  quantum distillation in the most occupied tubes. The inset of \textbf{j} shows
  the motion of the edges at $v_\text{max}$, as expected when the single particle quasimomentum
  distribution includes the midpoint of the band.
  \label{fig:2}}
\end{figure}

Figures~\ref{fig:2}\textbf{a}--\ref{fig:2}\textbf{c} show the evolution of the total atom
distribution (M1) for $V_0$=3$E_\text{rec}$, 4$E_\text{rec}$, and 5$E_\text{rec}$. These depths
correspond to $U/J$ of 4.7, 6.8 and 9.6,\cite{walters_cotugno_13} respectively, in the one-band
Hubbard model,\cite{Fisher_Fisher_89,cazalilla_citro_review_11} where $U$ is the onsite repulsion
energy. All are characterized by a central core of atoms that steadily releases
atoms that tunnel away from the center. Figures~\ref{fig:2}\textbf{d}--\ref{fig:2}\textbf{f}
show the distributions of single atoms, which are derived from M1, M2 and M3 (see Supplementary
Information Figs. S1\textbf{a}--S1\textbf{f}). These curves show two dominant features. First,
the cores contain many single atoms. Second, the broader pedestals of the M1
distributions are composed nearly exclusively of single atom sites. The velocities of the leading
edges of the pedestals equal, to within $\sim$10\% systematic uncertainties, the calculated maximum
possible velocity ($v_\text{max}=2Ja/\hbar$, where $a=\lambda/2$ is the lattice spacing) for
single atoms tunneling in the lowest band (see insets in
Figures~\ref{fig:2}\textbf{a}--\ref{fig:2}\textbf{c}). These velocities start to decrease at
the end of the compensation range of the crossed dipole trap, ultimately Bragg scattering backwards;
we do not display data after atoms return to the core. Figures~\ref{fig:2}\textbf{g}--\ref{fig:2}\textbf{i}
show the distribution of doublons, derived from all three measurement types. The number of doublons
steadily decreases after the quench, but the widths of the doublon distributions barely change.
The triplon distributions (see Supplementary Information Figs. S1\textbf{g}-- S1\textbf{i})) have
the same width as the doublon distributions to within a ~10$\%$ uncertainty.

Figures~\ref{fig:2}\textbf{j}--\ref{fig:2}\textbf{l} (and Supplementary Information
Figs.~S1\textbf{j}--S1\textbf{l}) show the results of a Gutzwiller mean-field
calculation\cite{Jaksch_Zoller_02,jreissaty_carrasquilla_13} (see Methods), which simulates an
array of identical tubes with different
atom numbers as in the experiment, and discretizes the direction along the tubes\cite{Stoudenmire_Burke_12} so that
the standard single-band approximation is not used. The theory assumes an initial zero temperature
BEC. This means that finite temperature and quantum fluctuations due to the one-dimensional character of the system
are not taken into account. The initial distribution is thus an imperfect match to the experiment
(see Methods). The doublons initially expand farther in the theory than
in the experiment, presumably because of the long range initial phase coherence in the theory.
The early decrease in the theory's doublon density no doubt affects the details of the ensuing dynamics,
but qualitatively, the theory behaves like the experiment in all respects other than the shape and size
of the doublon distributions. For additional comparison to theory, see Supplementary Information.

%%%%%%%%%%%%%%%%%%%%%%%%%%%%%%%%%%%%%%%%%%%%%%%%%%%%%%%%%%%%%%%%%%%%%%%%%
% FIGURE 2
%%%%%%%%%%%%%%%%%%%%%%%%%%%%%%%%%%%%%%%%%%%%%%%%%%%%%%%%%%%%%%%%%%%%%%%%%
%%%%%%%%%%%%%%%%%%%%%%%%%%%%%%%%%%%%%%%%%%%%%%%%%%%%%%%%%%%%%%%%%%%%%%%%%
% END FIGURE 2
%%%%%%%%%%%%%%%%%%%%%%%%%%%%%%%%%%%%%%%%%%%%%%%%%%%%%%%%%%%%%%%%%%%%%%%%%

\begin{figure}
  \begin{center}
    \includegraphics[width=1\columnwidth]{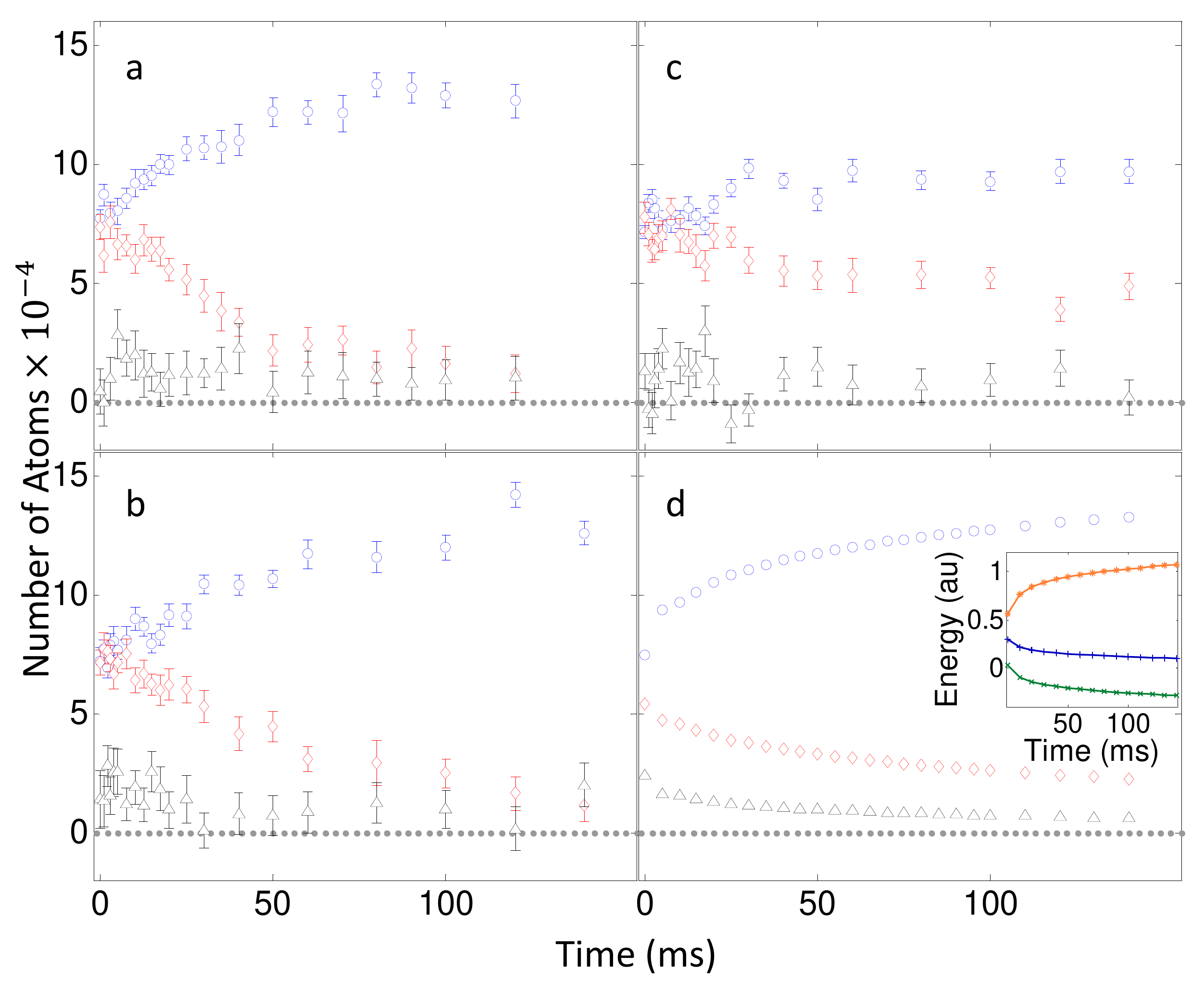}
  \end{center}
  \vspace*{-0.9cm}
  \caption{Time evolution of singlons, doublons and triplons. \textbf{a}. $V_0=3E_\text{rec}$ experiment.
  \textbf{b}. $V_0=4E_\text{rec}$ experiment. \textbf{c}. $V_0=5E_\text{rec}$ experiment. The error bars
  derive from the standard deviations determined from 8 separate measurements. When the curves in
  \textbf{a-c} are plotted together with a time axis rescaled by $J$ (see Supplementary Figure S2), they
  overlap well until the doublons reach their asymptotic number. \textbf{d}. $V_0=4E_\text{rec}$ theory.
  In all these subfigures the singlons are the blue circles, the doublons are the red diamonds, and the triplons are
  the black triangles. Inset to \textbf{d}: The interaction (blue pluses), kinetic (orange asterisks), and lattice potential
  (green crosses) energies vs.~$t_\text{ev}$ calculated for $V_0=4E_\text{rec}$.
   \label{fig:3}}
\end{figure}

Figures~\ref{fig:3}\textbf{a}--\ref{fig:3}\textbf{c} show the number of singlons, doublons and
triplons as a function of time, derived from the appropriate combinations of M1, M2, and M3. The
numbers of doublons and triplons drop steadily, with corresponding increases in the numbers of singlons.
The theory shows similar behavior (see Fig.~\ref{fig:3}\textbf{d}).
Although conservation of energy dictates that isolated doublons cannot dissociate for
$U/J\gtrsim4$,\cite{winkler_thalhammer_06} in a predominantly doublon sea the aforementioned
tunneling enhancement doubles this limit. Similarly,
one can show that a sea of singlons increases the limit by 50$\%$. That our doublons live in a bath
intermediate to these two seas explains why they dissociate, at least for $U/J\lesssim8$, and
why the dissociation rate decreases at long times (see especially Fig.~\ref{fig:3}\textbf{a} after 50 ms)
when the number of empty sites in the center
increases. Dissociation for $U/J=9.6$ ($V_0=5E_\text{rec}$) naively requires that energy be shared among more
singlons,\cite{stronmaier_greif_10,sensarma_pekker_10} but it might be that the one-band Hubbard
model calculation\cite{walters_cotugno_13} overestimates the effective value of $U/J$.
The latter interpretation is supported by the fact that, at short times,
the evolution of all the curves in Figs.~\ref{fig:2} and \ref{fig:3} are approximately self-similar
when the time axes are multiplied by $J$ (see Supplementary Information Figure S2), to within the small differences
in the doublon distribution widths discussed below. Though the physics is dominated by interacting
particle effects, marginal changes in site occupancy scale with $J$. Our mean field calculation
allows us to explicitly track the conversion of potential energy (interaction + lattice potential)
into kinetic energy (see inset) that results primarily from doublon dissolution.

%%%%%%%%%%%%%%%%%%%%%%%%%%%%%%%%%%%%%%%%%%%%%%%%%%%%%%%%%%%%%%%%%%%%%%%%%
% FIGURE 3
%%%%%%%%%%%%%%%%%%%%%%%%%%%%%%%%%%%%%%%%%%%%%%%%%%%%%%%%%%%%%%%%%%%%%%%%%
%%%%%%%%%%%%%%%%%%%%%%%%%%%%%%%%%%%%%%%%%%%%%%%%%%%%%%%%%%%%%%%%%%%%%%%%%
% END FIGURE 3
%%%%%%%%%%%%%%%%%%%%%%%%%%%%%%%%%%%%%%%%%%%%%%%%%%%%%%%%%%%%%%%%%%%%%%%%%

\begin{figure}
  \begin{center}
    \includegraphics[width=0.7\columnwidth]{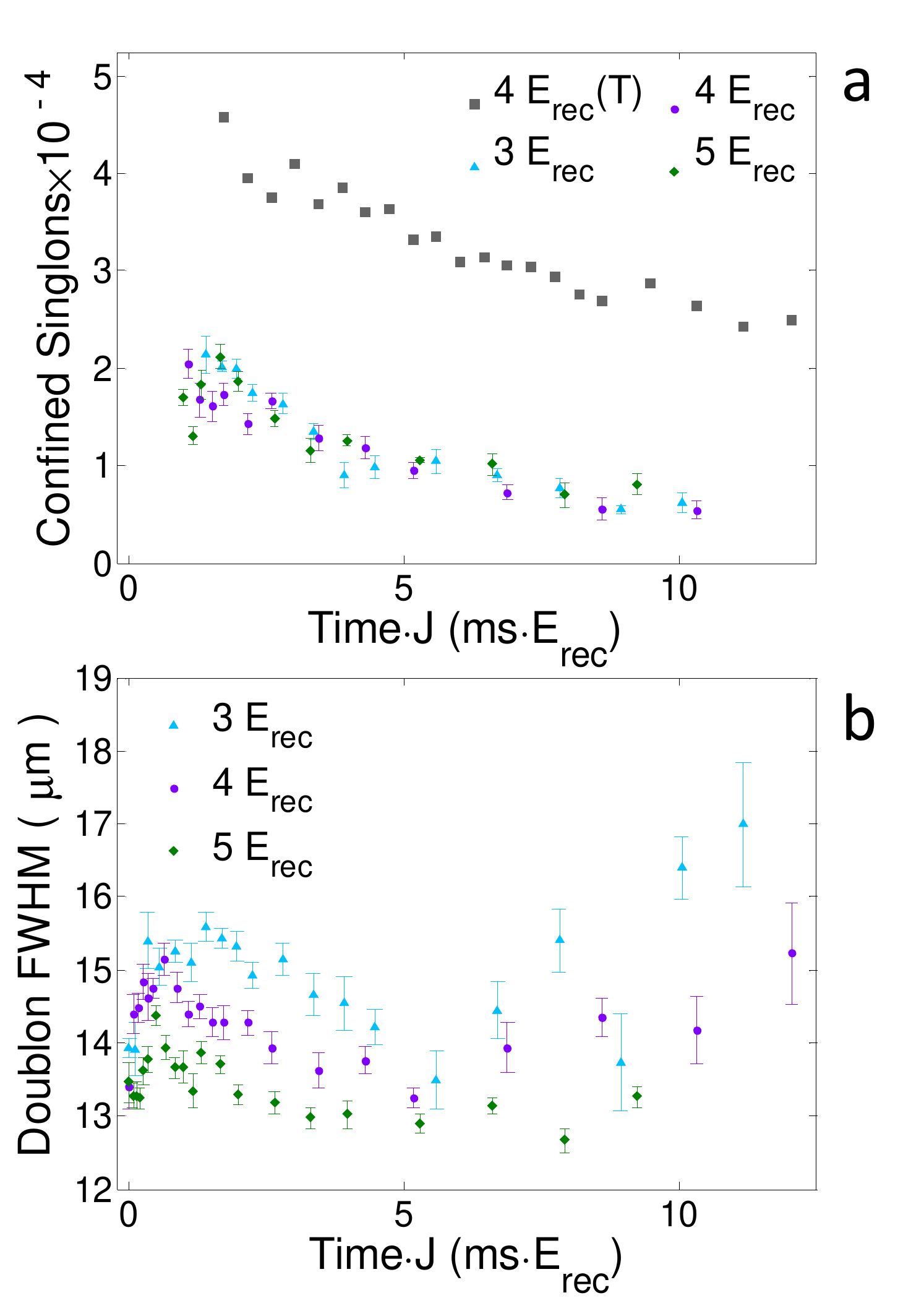}
  \end{center}
  \vspace*{-1cm}
  \caption{\textbf{a}. The number of singlons that are confined in the doublon sea vs the tunneling-rescaled
  evolution time. The triangles, circles, and diamonds are for $V_0$=3$E_\text{rec}$, 4$E_\text{rec}$ and
  5$E_\text{rec}$ respectively in the experiment. The squares are the theory results for $V_0$=4$E_\text{rec}$.
  See Methods for a description of how the data is derived from the singlon
  distributions.  \textbf{b}. The FWHM of the doublon distributions vs. tunneling-rescaled evolution time.
  (same labels as in \textbf{a}). The error bars in this figure derive from the standard deviation of
  8 measurements.
    \label{fig:4}}
\end{figure}

Quantum distillation is difficult to isolate at early times, since it occurs while initially unconfined
singlons are also leaving the central region and singlons are being created by dissolution.
But quantum distillation dominates at 5$E_\text{rec}$ (Fig.~\ref{fig:2}\textbf{f})
after 20 ms, by which time the doublon number is stable (see also Figs.~\ref{fig:2}\textbf{i} and
\ref{fig:3}\textbf{c}) and the unconfined singlons present a
locally flat background. The number of singlons confined in the doublon sea as a function of time is
plotted in Fig.~\ref{fig:4}\textbf{a} (see Methods). Its steady decrease is a clear signature of quantum
distillation, further supported by the fact that the rate scales with $J$ (see also Supplementary Information). At late times, the fraction of confined singlons
levels off at $\sim5\%$, showing long term vacancy confinement in the doublon sea. The
mean-field calculations also show some singlons initially leaving and others remaining indefinitely
(see Fig.~\ref{fig:3}\textbf{d}), but fewer singlons distill out in the calculations. This is expected
because the real 1D gas has a broader initial quasimomentum distribution. Thus in the calculation there
are more singlons with the lower energies that do not transmit out of the doublon sea.

In $J$-rescaled time, after doublons stop dissolving at 5$E_\text{rec}$ they are still dissolving at lower lattice depths. That the three sets of data points in Fig.~\ref{fig:4}\textbf{a} overlap means that extra doublon dissolution does not affect the central singlon number. This could be because the vast majority of singlons created when doublons dissolve have the right quasimomentum for immediate quantum distillation, and leave the center rapidly.

Further evidence of quantum distillation is given in Fig.~\ref{fig:4}\textbf{b}, which shows the
evolution of the full width at half maximum (FWHM) of the doublon distribution. There are three size
changing processes, each dominating for a time. The FWHMs increase during the first 10 to 20 ms because
doublons initially can expand into singly occupied sites and perhaps there is more doublon dissolution
in the middle (see Figures~\ref{fig:2}\textbf{g}--\ref{fig:2}\textbf{i}). The FWHMs then decrease due to
the mechanics of distillation, where escaping singlons move the last doublon one site inward. When the rate
of quantum distillation decreases (near 4 ms$\cdot$$E_\text{rec}$ in Fig.~\ref{fig:4}\textbf{a}) then as
long as the lattice depth is small

%%%%%%%%%%%%%%%%%%%%%%%%%%%%%%%%%%%%%%%%%%%%%%%%%%%%%%%%%%%%%%%%%%%%%%%%%
% FIGURE 4
%%%%%%%%%%%%%%%%%%%%%%%%%%%%%%%%%%%%%%%%%%%%%%%%%%%%%%%%%%%%%%%%%%%%%%%%%
%%%%%%%%%%%%%%%%%%%%%%%%%%%%%%%%%%%%%%%%%%%%%%%%%%%%%%%%%%%%%%%%%%%%%%%%%
% END FIGURE 4
%%%%%%%%%%%%%%%%%%%%%%%%%%%%%%%%%%%%%%%%%%%%%%%%%%%%%%%%%%%%%%%%%%%%%%%%%

\noindent enough, shrinking is overtaken by expansion. We suspect this to be
the result of higher order processes involving confined singlons that compromise the stability
of the edges of the doublon sea.\cite{petrosyan_schmidt_07} This unanticipated higher order effect,
undoubtedly absent in fermions and not present in the mean field theory results, further limits
the effectiveness of bosonic quantum distillation in producing low entropy blocks of doublons. 
The approximate stability or slight increase in the width of the doublon distribution implies 
that as time evolves, the number of empty sites among the doublons increases.

%%%%%%%%%%%%%%%%%%%%%%%%%%%%%%%%%%%%%%%%%%%%%%%%%%%%%%%%%%%%%%%%%%%%%%%%%%

% Conclusions

%%%%%%%%%%%%%%%%%%%%%%%%%%%%%%%%%%%%%%%%%%%%%%%%%%%%%%%%%%%%%%%%%%%%%%%%%%

Our work has concentrated on spatial distributions, which are local properties, but it
should also be possible to use related techniques like time-of-flight measurements to study non-local
properties, like quasimomentum
distributions and correlations. This simple lattice system, in which the doublon sea is open but
nonetheless settles to a stable steady state, can help address major open questions in quantum dynamics
such as how systems thermalize in the presence of particle losses.\cite{makotyn_klauss_14} Studying how
quantum correlations grow after the quench should give more general insight into how entanglement
spreads in quantum systems.\cite{cheneau2012} The fact that our mean-field treatment (only exact in
infinite dimensions) qualitatively captures the 1D dynamics, suggest that similar dynamics
occur in higher dimensions. Finally, an experimental implementation with fermions holds the promise
of producing superlatively low entropy doublon cores,\cite{fabian_manmana_09} which might allow the study
of hitherto inaccessible models of quantum magnetism \cite{balents_10} and high temperature
superconductivity.\cite{esslinger_review_10}

\section*{Methods}
\noindent \textbf{Experiment}\\
{\it Trapping:} We start with 2$\times10^5$ Bose condensed $^{87}$Rb atoms in the $F=1$, $m_F=1$ state in a crossed
dipole trap\cite{kinoshita_wenger_05} with 1.8 W per beam and 160 \SI{}{\micro\metre} beam waists. Gravity
is canceled by a magnetic field gradient. A blue-detuned
773.5 nm wavelength 3D optical lattice, made from two retroreflected horizontal 450 \SI{}{\micro\metre} waist
beams and a retroreflected vertical 700 \SI{}{\micro\metre} waist beam, is turned on in 14 ms
to a depth of $V_0$, after which the two horizontal lattice beam pairs are increased to their full depth
of 40$E_\text{rec}$ in 44 ms.  We find that the results of the experiment (including spatial distributions and
occupancy fractions) do not significantly change as long as the lattice turn-on times are 35 ms or longer.
That remains true if we wait for tens of ms in all the traps before starting the evolution.\\
{\it Flat Lattice:} Since for technical reasons the lattice waists are much larger than the dipole trap
waists, we can only create a flat lattice near the center of the trap. To fine tune the cancelation
of the two potentials, we start with trapped 1D gasses with no axial lattice, and choose the highest
crossed dipole beam intensity at which no atoms remain trapped in the central region. This gives a central
potential that is flat to within 0.08$E_\text{rec}$ over a length of 160 $\mu$m.\\
{\it Small bump:} The small bump on the left of the initial atom distributions in the experimental results
of Fig. 2 is due to spatial imperfections on the Yag beams that make up the confining crossed dipole trap,
where the atoms are trapped 3.4 Rayleigh lengths from the beam focus. After $t_\text{ev}$=0 the atoms evolve in
a much smoother potential, so only the initial distribution is affected by this issue, not the evolution.
The bump can serve as a feature, since it provides confirmatory evidence of the fraction of singlons that
remain confined (see, e.g., the long time curves in Fig. 2\textbf{f}).\\
{\it Determination of the trapped singlon fraction:} We analyze the singlon distributions (see Figures 2\textbf{d}--2\textbf{f}), by first determining
the FWHM of the doublons (see Figures 2\textbf{g}--2\textbf{i} and Figure 4\textbf{b})
at each time. We then measure the difference between the peak of the singlon distribution and its value at
the doublon FWHM positions. We assume that the confined peak height is twice that difference, and has the
doublon width. We do not use this procedure at very early times, before there is a discernible shoulder in
the singlon distribution, although the curves in Figure 4\textbf{a} change little if we add a few earlier points.
The assumption that the FWHMs of doublons and trapped singlons are the same is not exactly true for a bundle of
tubes. This procedure also assumes that the unconfined singlon distribution is approximately flat in the central
region, which is also not quite true. Our confidence in the reliability of this procedure is buttressed by the
universality of the curves in Figure 4\textbf{a}, which holds despite the systematic differences in the doublon
FWHMs at different $V_0$ (see Figure 4\textbf{b}).

\noindent \textbf{Theoretical calculations}\\
For the theoretical analysis of the expansion, we model a collection of independent
1D tubes. In order to account for possible higher band effects during the dynamics, which may be important
for the lowest lattice depths, we do not use the one-band approximation that results in the standard
Bose-Hubbard model. Instead, we discretize the space along the tubes by introducing an artificial
grid of spacing $\ell\ll\lambda$ (we take $\ell=0.05\lambda$, where $\lambda$ is the wavelength of the
optical lattice) to obtain a representation of the continuum in terms of an artificial one-dimensional
Bose-Hubbard model.\cite{Stoudenmire_Burke_12} The particles are then subject to the periodic potential
generated by the optical lattice and to the confining potentials generated by the crossed optical dipole
trap. We simulate arrays of up to $110\times110$ tubes, with each tube's length being $500\lambda$. We
choose the theory initial conditions so that the fraction of singlons in the simulation matches the experiment.
This is done by arbitrarily tuning the strength of the $t_\text{ev}<0$ axial and transverse trapping potentials, keeping their
ratio fixed to the experimental value. The parameters for $t_\text{ev}\geq0$ are those in the experimental setup,
except that the overall confining potential is set exactly to zero. The calculations are carried out within the
Gutzwiller mean-field approximation,\cite{Jaksch_Zoller_02,jreissaty_carrasquilla_13}
where the initial state is selected to be the ground state in the absence of tunneling between the
1D tubes. Our results are robust to further reduction of the value of $\ell$. For further details,
see Supplementary Information.

\begin{addendum}
\item We are indebted to Andrew Daley and Michael Fleischauer. This work was supported by NSF Grant No.
      PHYS-167830, the Army Research Office, and the Office of Naval Research (J.C. and M.R.).
      J.C. acknowledges support from the John Templeton Foundation. Research at Perimeter Institute
      is supported through Industry Canada and by the Province of Ontario through the Ministry of
      Research \& Innovation.
\end{addendum}

\bibliographystyle{naturemag}

\begin{thebibliography}{10}
\expandafter\ifx\csname url\endcsname\relax
  \def\url#1{\texttt{#1}}\fi
\expandafter\ifx\csname urlprefix\endcsname\relax\def\urlprefix{URL }\fi
\providecommand{\bibinfo}[2]{#2}
\providecommand{\eprint}[2][]{\url{#2}}

\bibitem{kinoshita_wenger_06}
\bibinfo{author}{Kinoshita, T.}, \bibinfo{author}{Wenger, T.} \&
  \bibinfo{author}{Weiss, D.~S.}
\newblock \bibinfo{title}{A quantum {Newton's} cradle}.
\newblock \emph{\bibinfo{journal}{Nature}} \textbf{\bibinfo{volume}{440-443}},
  \bibinfo{pages}{900--903} (\bibinfo{year}{2006}).

\bibitem{winkler_thalhammer_06}
\bibinfo{author}{Winkler, K.} \emph{et~al.}
\newblock \bibinfo{title}{Repulsively bound atom pairs in an optical lattice}.
\newblock \emph{\bibinfo{journal}{Nature}} \textbf{\bibinfo{volume}{441}},
  \bibinfo{pages}{853--856} (\bibinfo{year}{2006}).

\bibitem{ronzheimer_schreiber_13}
\bibinfo{author}{Ronzheimer, J.~P.} \emph{et~al.}
\newblock \bibinfo{title}{Expansion dynamics of interacting bosons in
  homogeneous lattices in one and two dimensions}.
\newblock \emph{\bibinfo{journal}{Phys. Rev. Lett.}}
  \textbf{\bibinfo{volume}{110}}, \bibinfo{pages}{205301}
  (\bibinfo{year}{2013}).

\bibitem{theis_thalhammer_04}
\bibinfo{author}{Theis, M.} \emph{et~al.}
\newblock \bibinfo{title}{Tuning the scattering length with an optically
  induced {F}eshbach resonance}.
\newblock \emph{\bibinfo{journal}{Phys. Rev. Lett.}}
  \textbf{\bibinfo{volume}{93}}, \bibinfo{pages}{123001}
  (\bibinfo{year}{2004}).

\bibitem{kinoshita_wenger_04}
\bibinfo{author}{Kinoshita, T.}, \bibinfo{author}{Wenger, T.} \&
  \bibinfo{author}{Weiss, D.~S.}
\newblock \bibinfo{title}{Observation of a one-dimensional {Tonks-Girardeau}
  gas}.
\newblock \emph{\bibinfo{journal}{Science}} \textbf{\bibinfo{volume}{305}},
  \bibinfo{pages}{1125--1128} (\bibinfo{year}{2004}).

\bibitem{fabian_manmana_09}
\bibinfo{author}{Heidrich-Meisner, F.} \emph{et~al.}
\newblock \bibinfo{title}{Quantum distillation: Dynamical generation of
  low-entropy states of strongly correlated fermions in an optical lattice}.
\newblock \emph{\bibinfo{journal}{Phys. Rev. A}} \textbf{\bibinfo{volume}{80}},
  \bibinfo{pages}{041603(R)} (\bibinfo{year}{2009}).

\bibitem{muth_petrosyan_12}
\bibinfo{author}{Muth, D.}, \bibinfo{author}{Petrosyan, D.} \&
  \bibinfo{author}{Fleischhauer, M.}
\newblock \bibinfo{title}{Dynamics and evaporation of defects in
  {M}ott-insulating clusters of boson pairs}.
\newblock \emph{\bibinfo{journal}{Phys. Rev. A}} \textbf{\bibinfo{volume}{85}},
  \bibinfo{pages}{013615} (\bibinfo{year}{2012}).

\bibitem{makotyn_klauss_14}
\bibinfo{author}{Makotyn, P.}, \bibinfo{author}{Klauss, C.~E.},
  \bibinfo{author}{Goldberger, D.~L.}, \bibinfo{author}{Cornell, E.~A.} \&
  \bibinfo{author}{Jin, D.~S.}
\newblock \bibinfo{title}{Universal dynamics of a degenerate unitary {B}ose
  gas}.
\newblock \emph{\bibinfo{journal}{Nature Phys.}} \textbf{\bibinfo{volume}{10}},
  \bibinfo{pages}{116--119} (\bibinfo{year}{2014}).

\bibitem{cheneau2012}
\bibinfo{author}{Cheneau, M.} \emph{et~al.}
\newblock \bibinfo{title}{Light-cone-like spreading of correlations in a
  quantum many-body system}.
\newblock \emph{\bibinfo{journal}{Nature}} \textbf{\bibinfo{volume}{481}},
  \bibinfo{pages}{484--487} (\bibinfo{year}{2012}).

\bibitem{Daley_Zoller_12}
\bibinfo{author}{Daley, A.~J.}, \bibinfo{author}{Pichler, H.},
  \bibinfo{author}{Schachenmayer, J.} \& \bibinfo{author}{Zoller, P.}
\newblock \bibinfo{title}{Measuring entanglement growth in quench dynamics of
  bosons in an optical lattice}.
\newblock \emph{\bibinfo{journal}{Phys. Rev. Lett.}}
  \textbf{\bibinfo{volume}{109}}, \bibinfo{pages}{020505}
  (\bibinfo{year}{2012}).

\bibitem{rigol_muramatsu_04}
\bibinfo{author}{Rigol, M.} \& \bibinfo{author}{Muramatsu, A.}
\newblock \bibinfo{title}{Emergence of quasicondensates of hard-core bosons at
  finite momentum}.
\newblock \emph{\bibinfo{journal}{Phys. Rev. Lett.}}
  \textbf{\bibinfo{volume}{93}}, \bibinfo{pages}{230404}
  (\bibinfo{year}{2004}).

\bibitem{rodriguez_manmana_06}
\bibinfo{author}{Rodriguez, K.}, \bibinfo{author}{Manmana, S.~R.},
  \bibinfo{author}{Rigol, M.}, \bibinfo{author}{Noack, R.~M.} \&
  \bibinfo{author}{Muramatsu, A.}
\newblock \bibinfo{title}{Coherent matter waves emerging from
  {Mott}-insulators}.
\newblock \emph{\bibinfo{journal}{New J. Phys.}} \textbf{\bibinfo{volume}{8}},
  \bibinfo{pages}{169} (\bibinfo{year}{2006}).

\bibitem{rigol_muramatsu_05}
\bibinfo{author}{Rigol, M.} \& \bibinfo{author}{Muramatsu, A.}
\newblock \bibinfo{title}{Fermionization in an expanding 1d gas of hard-core
  bosons}.
\newblock \emph{\bibinfo{journal}{Phys. Rev. Lett.}}
  \textbf{\bibinfo{volume}{94}}, \bibinfo{pages}{240403}
  (\bibinfo{year}{2005}).

\bibitem{minguzzi_gangardt_05}
\bibinfo{author}{Minguzzi, A.} \& \bibinfo{author}{Gangardt, D.~M.}
\newblock \bibinfo{title}{Exact coherent states of a harmonically confined
  {Tonks-Girardeau} gas}.
\newblock \emph{\bibinfo{journal}{Phys. Rev. Lett.}}
  \textbf{\bibinfo{volume}{94}}, \bibinfo{pages}{240404}
  (\bibinfo{year}{2005}).

\bibitem{vidmar_langer_13}
\bibinfo{author}{Vidmar, L.} \emph{et~al.}
\newblock \bibinfo{title}{Sudden expansion of {M}ott insulators in one
  dimension}.
\newblock \emph{\bibinfo{journal}{Phys. Rev. B}} \textbf{\bibinfo{volume}{88}},
  \bibinfo{pages}{235117} (\bibinfo{year}{2013}).

\bibitem{jreissaty_carrasquilla_13}
\bibinfo{author}{Jreissaty, A.}, \bibinfo{author}{Carrasquilla, J.} \&
  \bibinfo{author}{Rigol, M.}
\newblock \bibinfo{title}{Self-trapping in the two-dimensional {Bose-Hubbard}
  model}.
\newblock \emph{\bibinfo{journal}{Phys. Rev. A}} \textbf{\bibinfo{volume}{88}},
  \bibinfo{pages}{031606} (\bibinfo{year}{2013}).

\bibitem{rigol_batrouni_09}
\bibinfo{author}{Rigol, M.}, \bibinfo{author}{Batrouni, G.~G.},
  \bibinfo{author}{Rousseau, V.~G.} \& \bibinfo{author}{Scalettar, R.~T.}
\newblock \bibinfo{title}{State diagrams for harmonically trapped bosons in
  optical lattices}.
\newblock \emph{\bibinfo{journal}{Phys. Rev. A}} \textbf{\bibinfo{volume}{79}},
  \bibinfo{pages}{053605} (\bibinfo{year}{2009}).

\bibitem{burt_ghrist_97}
\bibinfo{author}{Burt, E.~A.} \emph{et~al.}
\newblock \bibinfo{title}{Coherence, correlations, and collisions: What one
  learns about bose-einstein condensates from their decay}.
\newblock \emph{\bibinfo{journal}{Phys. Rev. Lett.}}
  \textbf{\bibinfo{volume}{79}}, \bibinfo{pages}{337--340}
  (\bibinfo{year}{1997}).

\bibitem{tolra_ohara_04}
\bibinfo{author}{Tolra, B.~L.} \emph{et~al.}
\newblock \bibinfo{title}{Observation of reduced three-body recombination in a
  correlated 1d degenerate bose gas}.
\newblock \emph{\bibinfo{journal}{Phys. Rev. Lett.}}
  \textbf{\bibinfo{volume}{92}}, \bibinfo{pages}{190401}
  (\bibinfo{year}{2004}).

\bibitem{walters_cotugno_13}
\bibinfo{author}{Walters, R.}, \bibinfo{author}{Cotugno, G.},
  \bibinfo{author}{Johnson, T.~H.}, \bibinfo{author}{Clark, S.~R.} \&
  \bibinfo{author}{Jaksch, D.}
\newblock \bibinfo{title}{{\it Ab initio} derivation of {H}ubbard models for
  cold atoms in optical lattices}.
\newblock \emph{\bibinfo{journal}{Phys. Rev. A}} \textbf{\bibinfo{volume}{87}},
  \bibinfo{pages}{043613} (\bibinfo{year}{2013}).

\bibitem{Fisher_Fisher_89}
\bibinfo{author}{Fisher, M. P.~A.}, \bibinfo{author}{Weichman, P.~B.},
  \bibinfo{author}{Grinstein, G.} \& \bibinfo{author}{Fisher, D.~S.}
\newblock \bibinfo{title}{Boson localization and the superfluid-insulator
  transition}.
\newblock \emph{\bibinfo{journal}{Phys. Rev. B}} \textbf{\bibinfo{volume}{40}},
  \bibinfo{pages}{546--570} (\bibinfo{year}{1989}).

\bibitem{cazalilla_citro_review_11}
\bibinfo{author}{Cazalilla, M.~A.}, \bibinfo{author}{Citro, R.},
  \bibinfo{author}{Giamarchi, T.}, \bibinfo{author}{Orignac, E.} \&
  \bibinfo{author}{Rigol, M.}
\newblock \bibinfo{title}{One dimensional bosons: From condensed matter systems
  to ultracold gases}.
\newblock \emph{\bibinfo{journal}{Rev. Mod. Phys.}}
  \textbf{\bibinfo{volume}{83}}, \bibinfo{pages}{1405--1466}
  (\bibinfo{year}{2011}).

\bibitem{Jaksch_Zoller_02}
\bibinfo{author}{Jaksch, D.}, \bibinfo{author}{Venturi, V.},
  \bibinfo{author}{Cirac, J.~I.}, \bibinfo{author}{Williams, C.~J.} \&
  \bibinfo{author}{Zoller, P.}
\newblock \bibinfo{title}{Creation of a molecular condensate by dynamically
  melting a {M}ott insulator}.
\newblock \emph{\bibinfo{journal}{Phys. Rev. Lett.}}
  \textbf{\bibinfo{volume}{89}}, \bibinfo{pages}{040402}
  (\bibinfo{year}{2002}).

\bibitem{Stoudenmire_Burke_12}
\bibinfo{author}{Stoudenmire, E.~M.}, \bibinfo{author}{Wagner, L.~O.},
  \bibinfo{author}{White, S.~R.} \& \bibinfo{author}{Burke, K.}
\newblock \bibinfo{title}{One-dimensional continuum electronic structure with
  the density-matrix renormalization group and its implications for
  density-functional theory}.
\newblock \emph{\bibinfo{journal}{Phys. Rev. Lett.}}
  \textbf{\bibinfo{volume}{109}}, \bibinfo{pages}{056402}
  (\bibinfo{year}{2012}).

\bibitem{stronmaier_greif_10}
\bibinfo{author}{Strohmaier, N.} \emph{et~al.}
\newblock \bibinfo{title}{Observation of elastic doublon decay in the
  {Fermi-Hubbard} model}.
\newblock \emph{\bibinfo{journal}{Phys. Rev. Lett.}}
  \textbf{\bibinfo{volume}{104}}, \bibinfo{pages}{080401}
  (\bibinfo{year}{2010}).

\bibitem{sensarma_pekker_10}
\bibinfo{author}{Sensarma, R.} \emph{et~al.}
\newblock \bibinfo{title}{Lifetime of double occupancies in the {Fermi-Hubbard}
  model}.
\newblock \emph{\bibinfo{journal}{Phys. Rev. B}} \textbf{\bibinfo{volume}{82}},
  \bibinfo{pages}{224302} (\bibinfo{year}{2010}).

\bibitem{petrosyan_schmidt_07}
\bibinfo{author}{Petrosyan, D.}, \bibinfo{author}{Schmidt, B.},
  \bibinfo{author}{Anglin, J.~R.} \& \bibinfo{author}{Fleischhauer, M.}
\newblock \bibinfo{title}{Quantum liquid of repulsively bound pairs of
  particles in a lattice}.
\newblock \emph{\bibinfo{journal}{Phys. Rev. A}} \textbf{\bibinfo{volume}{76}},
  \bibinfo{pages}{033606} (\bibinfo{year}{2007}).

\bibitem{balents_10}
\bibinfo{author}{Balents, L.}
\newblock \bibinfo{title}{Spin liquids in frustrated magnets}.
\newblock \emph{\bibinfo{journal}{Nature}} \textbf{\bibinfo{volume}{464}},
  \bibinfo{pages}{199--208} (\bibinfo{year}{2010}).

\bibitem{esslinger_review_10}
\bibinfo{author}{Esslinger, T.}
\newblock \bibinfo{title}{{Fermi-Hubbard} physics with atoms in an optical
  lattice}.
\newblock \emph{\bibinfo{journal}{Annual Review of Condensed Matter Physics}}
  \textbf{\bibinfo{volume}{1}}, \bibinfo{pages}{129--152}
  (\bibinfo{year}{2010}).

\bibitem{kinoshita_wenger_05}
\bibinfo{author}{Kinoshita, T.}, \bibinfo{author}{Wenger, T.} \&
  \bibinfo{author}{Weiss, D.~S.}
\newblock \bibinfo{title}{All-optical {Bose-Einstein} condensation using a
  compressible crossed dipole trap}.
\newblock \emph{\bibinfo{journal}{Phys. Rev. A}} \textbf{\bibinfo{volume}{71}},
  \bibinfo{pages}{011602} (\bibinfo{year}{2005}).

\end{thebibliography}

%%%%%%%%%%%%%%%%%%%%%%%%%%%%%%%%%%%%%%%%%%%%%%%%%%%%%%%%%%%%%%%%%%%%%%%%%%%%
%%%%%%%%%%%%%%%%%%%%%%%%%%%%%%%%%%%%%%%%%%%%%%%%%%%%%%%%%%%%%%%%%%%%%%%%%%%%



\section*{Supplementary material}

# Supplementary information: Quantum distillation and confinement of vacancies in a doublon sea

Lin Xia[1], Laura A. Zundel[1], Juan Carrasquilla[1,2], Aaron Reinhard[1], Joshua M. Wilson[1], Marcos Rigol[1], and David S. Weiss[1]

[1]Department of Physics, The Pennsylvania State University, University Park, PA 16802, USA
[2]Perimeter Institute for Theoretical Physics, Waterloo, Ontario, Canada N2L 2Y5

**Experiments**

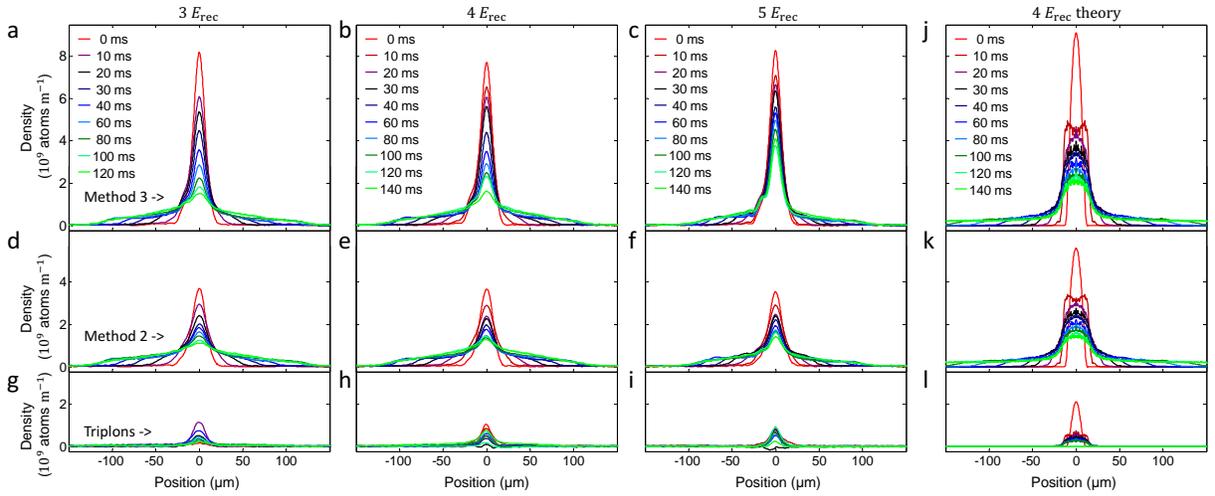

**Figure S1  Spatial dynamics in the flat lattice. a-c**. M3 measurements of atom distributions after three-body inelastic collisions have emptied triplon sites and most atoms from more highly occupied sites. **d-f**. M2 measurements of atom distributions after photoassociation. These distributions are approximately the same as the distributions of singlons. **g-i**. The distributions of triplons. These are derived from combining the raw measurements according to the formula M1-M3. Note that all the experimental plots (**a-i**) have the same vertical scale, and results are reported at successive $t_{ev}$ at $V_0$=3$E_{rec}$ (panels in the first column), 4$E_{rec}$ (panels in the second column), and 5$E_{rec}$ (panels in the third column). For a discussion of the small asymmetry in these figures, see Methods. **j-l.** Gutzwiller mean-field theory results for $V_0 = 4E_{rec}$ for M3, M1, and triplons, respectively.

1

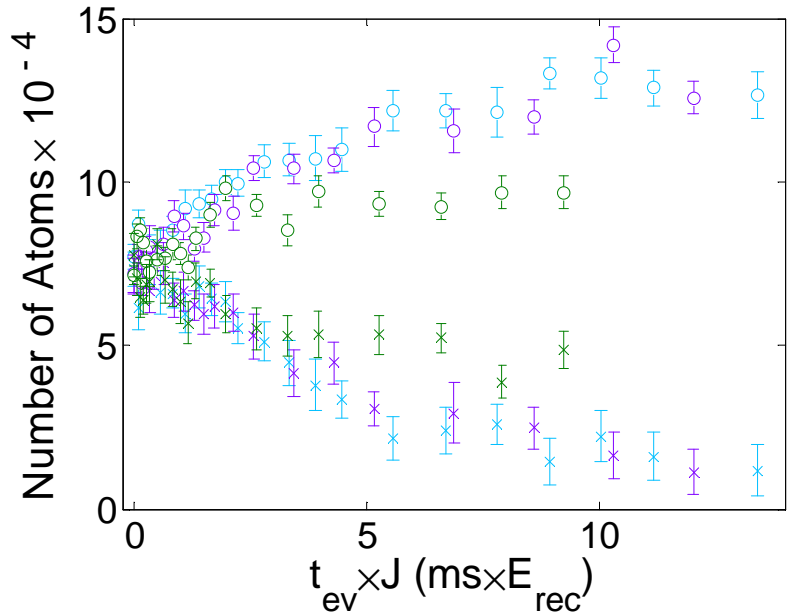

**Figure S2  The time-rescaled loss of doublons.** The experimental data from Fig. 3a-3b is replotted with the time axis scaled by J. The doublon populations are denoted by ×'s, and the singlon populations by open circles. The blue, purple and green symbols are from $V_0 = 3E_{rec}$, $4E_{rec}$, and $5E_{rec}$ respectively. The curves are approximately self-similar up until ∼3ms-$E_{rec}$, which suggests that doublon decay is predominantly a first-order process. The largest difference in the curves is that when the lattice is deeper, the doublons stop decaying at an earlier time, leaving a larger asymptotic doublon population.

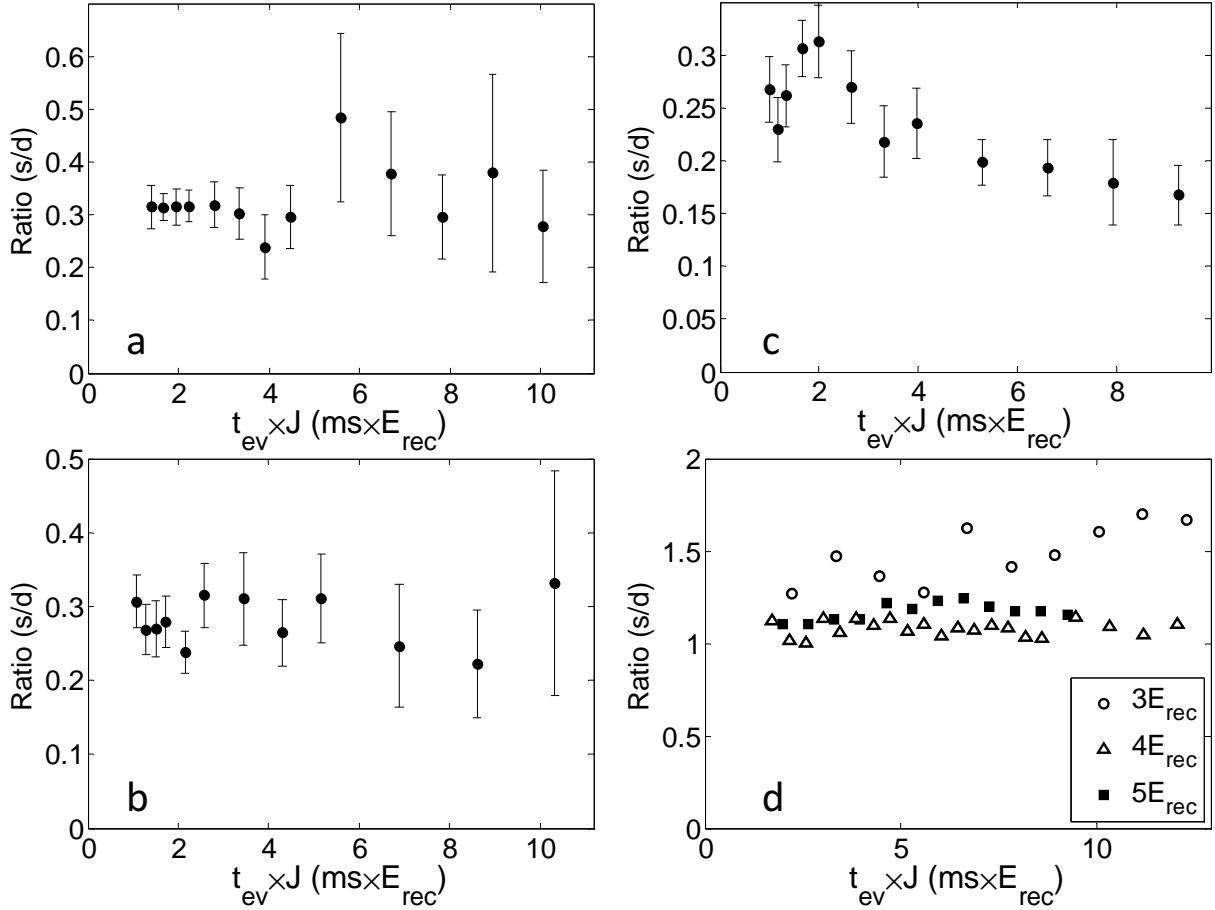

**Figure S3  The ratio of trapped singlons to doublons. a.** $3E_{rec}$ experimental data, derived from data in Figs. 3a and 4a. **b.** $4E_{rec}$ experimental data, derived from data in Figs. 3b and 4a. **c.** $5E_{rec}$ experimental data, derived from data in Figs. 3c and 4a. **d.** Theoretical results for the ratio of trapped singlons to doublons at the three lattice depths, derived from Gutzwiller mean field calculations in the same way as the corresponding results are derived from the experimental data. The downward trend in **c** is the most direct signature of quantum distillation, since the number of trapped singlons decreases more quickly than the number of doublon atoms. Because at the lower lattice depths the number of doublons continues to decrease by dissolution into singlons, the fact that the ratios in **a** and **b** are approximately constant also implies that quantum distillation is occurring there. Most of the new singlons from doublon dissolution quantum distill out, along with enough of the old singlon to keep pace with the decreasing number of doublon atoms. The theory results here can be understood in a similar way, noting that at the higher two depths, there is both less doublon dissolution and less quantum distillation than in the experiment. We suspect that the dominant reason for quantitative disagreements between the experiment and the theory is the initial condition. The theory distributions are initially colder, which leads to an initial density-dropping expansion of all the atoms, and thus a lower central density (with more singlons and empty sites) throughout the rest of the evolution.

3

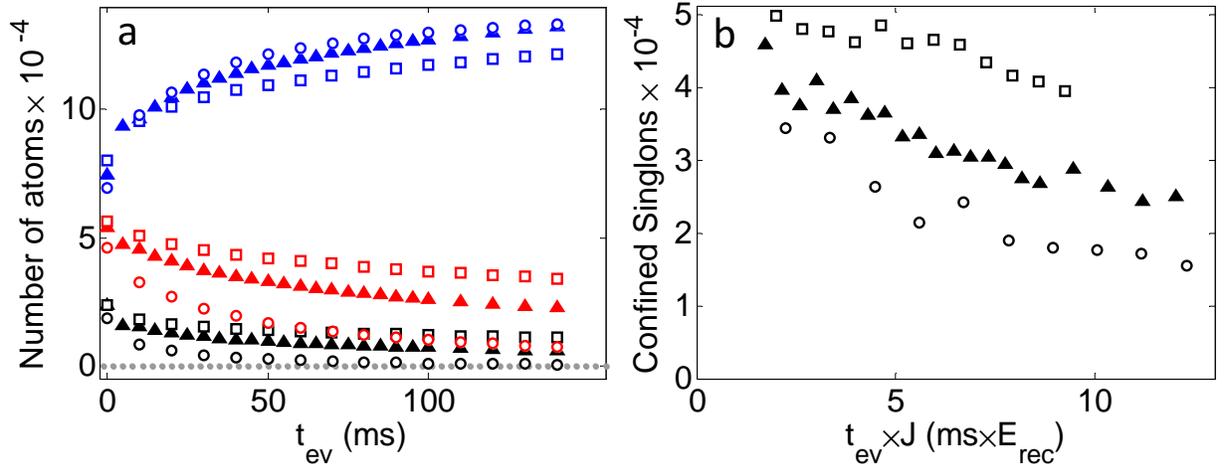

**Figure S4  Theoretical evolution curves. a.** The numbers of doublons (red), singlons (blue), and triplons (black) vs. evolution time. The circles are for $V = 3E_{\rm rec}$, the triangles are for $V = 4E_{\rm rec}$, and the squares are for $V = 5E_{\rm rec}$. **b.** The number of confined singlons vs. time, with data labels as in **a**. As with the theory shown in the body of the paper, there are strong qualitative similarities with the experimental data. Comparison of **a** to Fig. 3 suggests that the strong difference in the extent of doublon dissolution between $4E_{\rm rec}$ and $5E_{\rm rec}$ in the experiment is shifted to between $3E_{\rm rec}$ and $4E_{\rm rec}$ in the theory. As with the other quantitative disagreements we see, we suspect the initial conditions, which give rise to more initial expansion at the deeper lattice depths. The curves in **b** do not overlap each other like those in Fig. 4b. Unlike in the experiment, where the initial spatial distributions do not significantly vary with lattice depth, they do in the theory calculations.

4

**Theory**

We assume that the system under consideration can be described in terms of particles moving in an array of independent 1D tubes. This is justified because the tunneling amplitude between adjacent tubes is exponentially small in the strength of the optical lattice defining the array, which in the experiment is $V_{\text{transverse}}/E_{\text{rec}} = 40$. The Hamiltonian in each 1D tube is

$$\mathcal{H} = \int dz \psi^\dagger(z) \left[ -\frac{\hbar^2}{2m} \frac{d^2}{dz^2} + V(z) \right] \psi(z) + \frac{1}{2} \int dz \psi^\dagger(z) \psi^\dagger(z') U(z-z') \psi(z') \psi(z)$$
$$- \mu \int dz \psi^\dagger(z) \psi(z), \qquad (1)$$

where the field operators satisfy $[\psi(z), \psi^\dagger(z')] = \delta(z-z')$. The particles in each tube are subjected to a local potential due to the optical lattice and to an axial confinement potential, i.e., $V(z) = V_0 \sin^2\left(\frac{2\pi}{\lambda} z\right) + \frac{1}{2} m \omega_z^2 z^2$. The optical lattice is characterized by its depth $V_0$ and wavelength $\lambda$. The axial confinement is characterized by the frequency $\omega_z$ and the mass $m$ of the Rb$^{87}$ atoms. It is further assumed that the particles interact via a short-range potential $U(z-z') = g_{1D} \delta(z-z')$.[1] The chemical potential $\mu$ controls the total number of particles in the tube. In order to study the Hamiltonian in Eq. (1), one can then map the problem onto a simplified one-band Bose-Hubbard model,[2] which is a good approximation when the lattice is deep enough so that multiband effects can be neglected. Alternatively, it is possible to account for such effects, that are potentially relevant for the lowest lattice depths considered in the experiments, by introducing a grid spacing $\ell = \lambda/T$ to obtain the following "artificial" lattice Hamiltonian[3]

$$\mathcal{H}_{\text{BH}} = -t_h \sum_i \left( b_i^\dagger b_{i+1} + \text{h.c} \right) + \frac{U_b}{2} \sum_i n_i(n_i - 1) + V_0 \sum_i \sin^2\left(\frac{2\pi}{T} i\right) n_i + \Omega_z \sum_i i^2 n_i$$
$$- \tilde{\mu} \sum_i n_i, \qquad (2)$$

where we have identified $b_i^\dagger = \psi^\dagger(i\ell)/\sqrt{\ell}$. Here, the hopping matrix element $t_h = E_{\text{rec}} \left(\frac{T}{2\pi}\right)^2$, the interaction $U_b = g_{1D} \frac{T}{\lambda}$, and the amplitude of the axial trapping potential is given by $\Omega_z = \frac{1}{2} m \omega_z^2 \left(\frac{\lambda}{T}\right)^2$. Notice that the original Hamiltonian Eq. (1) is recovered in the limit $\mathcal{H} = \lim_{\ell \to 0} \mathcal{H}_{\text{BH}}$. In our simulations, however, we consider a finite $\ell = 0.05\lambda$ and ensure that the results are robust upon further increase of $T$. The coupling strength $g_{1D}$ is estimated following Ref. 1 based on the parameters of the experimental setup. The effective chemical potential of each independent tube $\tilde{\mu} = \mu - \left(\frac{T}{\lambda}\right)^2$ is obtained by adding an energy offset due to the transverse parabolic trapping potential (which is characterized by its frequency $\omega_{xy}$) to the overall chemical potential of the system. The latter is, in turn, fixed by total number of particles in the experiments. We simulate arrays of up to $110 \times 110$ tubes, each tube being $1000\ell$ long. The axial size of the system is chosen so that the expanding particles never reach the edges of the tubes in the time scales of the simulation. The initial state ($t_{\text{ev}} = 0$) is assumed to be the ground state in the absence of tunneling between the 1D tubes. We tune the $t_{\text{ev}} = 0$ axial and transverse trapping potentials

5

(using the experimental value for their ratio) so that the initial fraction of singlons in the simulation matches the experiment. The lattice parameters are those of the experiment. For $t_{\text{ev}} \geq 0$ the overall confining potential is set to zero exactly. All calculations are carried out using a Gutzwiller mean-field approach detailed in Refs. 4, 5.

The parameters $U/J$ that correspond to the one-band Bose-Hubbard model discussed in the main text were obtained using maximally-localized generalized Wannier states[6] through the software available in Ref. 7.

1. Olshanii, M. Atomic scattering in the presence of an external confinement and a gas of impenetrable bosons. *Phys. Rev. Lett.* **81**, 938–941 (1998).

2. Fisher, M. P. A., Weichman, P. B., Grinstein, G. & Fisher, D. S. Boson localization and the superfluid-insulator transition. *Phys. Rev. B* **40**, 546–570 (1989).

3. Stoudenmire, E. M., Wagner, L. O., White, S. R. & Burke, K. One-dimensional continuum electronic structure with the density-matrix renormalization group and its implications for density-functional theory. *Phys. Rev. Lett.* **109**, 056402 (2012).

4. Jaksch, D., Venturi, V., Cirac, J. I., Williams, C. J. & Zoller, P. Creation of a molecular condensate by dynamically melting a Mott insulator. *Phys. Rev. Lett.* **89**, 040402 (2002).

5. Jreissaty, A., Carrasquilla, J. & Rigol, M. Self-trapping in the two-dimensional Bose-Hubbard model. *Phys. Rev. A* **88**, 031606 (2013).

6. Walters, R., Cotugno, G., Johnson, T. H., Clark, S. R. & Jaksch, D. *Ab initio* derivation of Hubbard models for cold atoms in optical lattices. *Phys. Rev. A* **87**, 043613 (2013).

7. Johnson, T., Clark, S., & Cotugno, G. Wannier states for optical lattices (2013). URL http://ccpforge.cse.rl.ac.uk/gf/project/mlgws/.



\end{document}